\documentclass[prd,preprintnumbers,aps]{revtex4}
\usepackage{graphicx}
\usepackage{dcolumn}
\usepackage{bm}
\usepackage{epsfig,amsmath}
\usepackage{amssymb}
\usepackage{subfigure}
\usepackage{float}
\usepackage[usenames,dvipsnames]{color}
\usepackage[pagebackref=false, colorlinks=true]{hyperref}
\definecolor{redish}{rgb}{0.7,0.2,0.0}  
\definecolor{bluish}{rgb}{0.2,0.5,0.8}

\hypersetup{linkcolor=redish,          
                  citecolor=blue,        
                  filecolor=magenta,      
                  urlcolor=bluish}          

\DeclareFontFamily{U}{rsfs}{}         
\DeclareFontShape{U}{rsfs}{m}{n}{<5> rsfs5 <6><7> rsfs7          %
  <8><9><10><10.95><12><14.4><17.28><20.74><24.88> rsfs10}{}     %
\DeclareMathAlphabet{\mathfs}{U}{rsfs}{m}{n}

\def \O{\Omega}

\def \f{\frac}

\def \O{\Omega}

\def \p{\partial}
\def \d{\Delta}

\def \th{\theta}

\begin{document}
\title{Does the gravitomagnetic monopole exist? A clue from a black hole x-ray binary}

\author{Chandrachur Chakraborty}
\email{chandra@pku.edu.cn}
\affiliation{Kavli Insttute for Astronomy and Astrophysics, 
Peking University, Beijing 100871, China}
\author{Sudip Bhattacharyya}
\email{sudip@tifr.res.in}
\affiliation{Department of Astronomy and Astrophysics,
Tata Institute of Fundamental Research, Mumbai 400005, India}

\begin{abstract}
The gravitomagnetic monopole is the proposed gravitational analogue of Dirac's 
magnetic monopole.
However, an observational evidence of this aspect of fundamental physics was elusive.
Here, we employ a technique involving three primary X-ray observational methods used 
to measure a black hole spin to search for the gravitomagnetic monopole.
These independent methods give significantly
different spin values for an accreting black hole. We demonstrate 
that the inclusion of one extra parameter due to the gravitomagnetic monopole not only makes the spin
and other parameter values inferred from the three methods 
consistent with each other but also makes the inferred black hole mass consistent
with an independently measured value. We argue that this first indication of the
gravitomagnetic monopole, within our paradigm, is not a result of fine tuning.
\end{abstract}

\maketitle 

\section{Introduction}

The gravitational analogue of Dirac's magnetic monopole \cite{dirac, saha}
is known as the gravitomagnetic monopole \cite{zee}, which, if detected, 
can open a new area of research in physics. Historically, Newmann et al. discovered
a stationary and spherically symmetric exact solution (now known as the NUT solution) 
of Einstein equation,
that contains the gravitomagnetic monopole or the so-called NUT (Newman, Unti and
Tamburino \cite{nut}) parameter \cite{ahm, bini}. 
Note that Einstein-Hilbert action requires no modification \cite{rs2} to accommodate the gravitomagnetic monopole. 
Demianski and Newman found that the NUT spacetime is produced by a `dual mass' \cite{dn} or the 
gravitomagnetic charge/monopole. Bonnor \cite{bon} physically interpreted it 
as `a linear source of pure angular momentum' \cite{dow,rs}, i.e., `a massless rotating rod'.
Moreover, the NUT spacetime is free of curvature singularities \cite{rs2},
and the mass (or the so-called gravitoelectric charge) quantization \cite{zee} is possible
due to the presence of the gravitomagnetic charge, which is a general feature \cite{mp} 
of a spacetime with dual mass \cite{rs2, rs}. Therefore, gravitomagnetic monopole or NUT 
parameter is a fundamental aspect of physics.

While the existence of gravitomagnetic monopole is an exciting possibility, 
to the best of our knowledge, 
a serious effort to search for it among the astronomical objects 
has not been made so far. Lynden-Bell and Nouri-Zonoz \cite{lnbl}, 
who were possibly the first to motivate such an investigation, 
argued that the best place to look for the gravitomagnetic monopole is in the spectra of supernovae, 
quasars, or active galactic nuclei (see also \cite{kag}). 
But practical ways to detect the gravitomagnetic monopole in nature, if it exists, were not proposed.
In this paper, we demonstrate that X-ray observations of a black hole X-ray binary (BHXB),
i.e., an accreting stellar-mass collapsed object, can provide a way to 
detect a non-zero NUT parameter or the gravitomagnetic monopole.
This is because, while the spacetime of such a spinning collapsed object 
(a black hole, or even a naked singularity) is usually described with the Kerr metric 
\cite{kerr}, the Kerr geometry may naturally contain the NUT parameter along with 
the mass and the angular momentum, and be known  as the Kerr-Taub-NUT (KTN) spacetime \cite{ml},
which is geometrically a stationary, axisymmetric vacuum solution of Einstein equation, and 
reduces to the Kerr spacetime if the NUT parameter is zero. 
Therefore, identification of a collapsed object having the KTN 
spacetime with a non-zero NUT parameter can be ideal to establish the existence of the gravitomagnetic monopole.
In this paper, we demonstrate that X-ray observations of a black hole X-ray binary (BHXB),
i.e., an accreting stellar-mass collapsed object, can provide a way to infer and measure the NUT parameter.
Note that, while the collapsed object is usually thought to be a black hole, i.e., a singularity covered
by an event horizon, here we do not exclude the
possibility that it could also be a naked or uncovered singularity in some cases \cite{ckp}.

We search for the gravitomagnetic monopole using three independent X-ray
observational methods used to measure a black hole spin. We briefly discuss these methods
in Sec.\ref{s2}. In our study, we use fundamental frequencies, ISCO radius and
gravitational redshift. We derive and provide formulae for some of these quantities 
for various spacetimes in Sec.\ref{s3}. In Sec.\ref{s4}, we use these expressions
to explore the possibility of the non-zero NUT parameter in a BHXB : GRO J1655--40. 
The additional plausible solutions of our work 
are discussed in Sec.\ref{s5} and finally, we conclude in Sec.\ref{s6}.

\section{\label{s2}Materials and Methods}

Measurement of the NUT parameter can be done by combining several methods, which are used to 
measure the spin parameter (or Kerr parameter) $a/M$ of an accreting collapsed object.
Here, $a/M = J/M^2$, where $M$ and $J$ are the collapsed object mass and angular momentum 
respectively. Note that measuring $a/M$
can be very useful to probe the strong gravity regime and to characterize the collapsed object, 
and a significant effort in astronomy has been made for such measurements \cite{RemillardMcClintock2006, Miller2007, bs14}.
However, different methods to measure $a/M$ do not sometimes give consistent results,
which make these methods unreliable. In this paper, we demonstrate that these results can be consistent with each 
other, if we allow a non-zero NUT parameter value.

Some of the X-ray spectral and timing features, originating from the accreted matter
within a few gravitational radii of a collapsed object in a BHXB,
can be used to measure the spin parameter $a/M$ \cite{RemillardMcClintock2006, Miller2007, bs14}. 
There are two main spectral methods for $a/M$ estimation: (1) using broad relativistic 
iron K$\alpha$ spectral emission line \cite{Miller2007}, and (2) using continuum X-ray spectrum
\cite{RemillardMcClintock2006}. There is also a timing method 
based on the relativistic precession model (RPM) of quasi-periodic oscillations (QPOs) 
of X-ray intensity \cite{motta40}. We briefly discuss these methods below.

A broad relativistic iron K$\alpha$ spectral emission line in X-rays is observed
from many BHXBs, and such a fluorescent line is believed to originate from the
reflection of hard X-rays from the inner part of the geometrically thin accretion disk.
This intrinsically narrow iron line ($6.4-6.97$~keV)
is broadened, becomes asymmetric and shifted towards lower energies by physical effects,
such as Doppler effect, special relativistic beaming and gravitational redshift 
\cite{ReynoldsNowak2003, Miller2007}.
Note that it is primarily the extent of the red wing of the line that determines the observed constraint
on $a/M$ \cite{Reisetal2009}. This is because this red wing extent gives a measure of the
gravitational redshift at the disk inner edge radius $r_{\rm in}$ (as this redshift in the disk is 
the maximum at the inner edge), and for $r_{\rm in} = r_{\rm ISCO}$ ($r_{\rm ISCO}$ is the 
innermost stable circular orbit (ISCO) radius), the $a/M$ value can be inferred
for a prograde accretion disk in the Kerr spacetime (Eqs.~\ref{iscokerr} and \ref{zkerr}).

The modeling of the observed continuum X-ray spectrum 
can also be used to constrain $a/M$. In this method, the thermal spectral component from the
accretion disk is fit with a relativistic thin-disk model, and this gives a measure of the 
$r_{\rm in}$, if the source distance ($D$) and the accretion disk inclination angle ($i$) 
are independently measured \cite{FragosMcClintock2015}.
Then from a known $M$ value, $r_{\rm ISCO}/M$, and hence $a/M$, can be inferred assuming Kerr spacetime.

The QPO-based timing method uses three observed features to estimate $a/M$: (a) the upper high-frequency (HF) QPO, 
(b) the lower HFQPO, and (c) the type-C low-frequency (LF) QPO \cite{motta40}.
HFQPOs are rare, and they are observed in the frequency range of $\sim 40 - 450$~Hz \cite{Bellonietal2012}.
Type-C QPO is the most common LFQPO, and it is observed in the frequency range of 
$\sim 0.01 - 30$~Hz \cite{bs14}.
According to this method based on the relativistic precession model (RPM), 
which was first proposed for accreting neutron stars by \cite{StellaVietri1998,
StellaVietri1999}, the Type-C QPO frequency is identified with the LT precession frequency ($\nu_{\rm LT}$), 
and the upper and lower HFQPO frequencies are identified with the orbital frequency ($\nu_{\phi}$) and 
the periastron precession frequency ($\nu_{\rm per}$) respectively \cite{motta40}. 
For the Kerr spacetime, each of these three frequencies is a function of three parameters: 
$M$, $a/M$ and the radial coordinate $r_{\rm qpo}$ of the location of origin of these QPOs. Hence, 
the RPM method can provide not only the $a/M$ value, but also the values of $M$ and $r_{\rm qpo}$ \cite{motta40}.

So far the RPM method could be fully applied for one BHXB (GRO J1655--40), because, to the
best of our knowledge, the three above-mentioned QPOs could be simultaneously observed
only from this BHXB \cite{motta40}.
The mass of the collapsed object of GRO J1655--40 is either $(6.3\pm0.5)M_\odot$ \cite{Greeneetal2001}
or $(5.4\pm0.3)M_\odot$ \cite{beer} (it is not yet clear which one is more reliable; 
\cite{FragosMcClintock2015}). 
According to the RPM, the observed frequencies of the above mentioned three simultaneous QPOs imply
$\nu_{\phi} = 440$~Hz, $\nu_{\rm per} = 300$~Hz and $\nu_{\rm LT} = 17$~Hz for GRO J1655--40. Using these
frequencies, \cite{motta40} determined $a/M \approx 0.286\pm0.003$, $M \approx (5.31\pm0.07)M_\odot$
and $r_{\rm qpo} \approx (5.68\pm0.04)~M$.
Moreover, the inferred $M \approx 5.31M_\odot$ is consistent with 
an independently measured mass $(5.4\pm0.3)M_\odot$ \cite{beer}, which indicates the reliability of
the RPM method and the corresponding inferred parameter values for GRO J1655--40.

While such a QPO-based estimation of the $a/M$ value is model dependent, we would like to note that 
a recent observation of a variation of the relativistic iron line energy with the phase of
the Type-C QPO from the BHXB H1743-322 supports that this QPO is caused by
the LT precession (\cite{in16}; see also \cite{MillerHoman2005, Schnittmanetal2006}),
as considered in the RPM. Note that this may require a tilted inner disk, which has recently been
theoretically shown to be possible \cite{ChakrabortyBhattacharyya2017}.
Besides, while the RPM interpretation of HFQPOs is not unique, the reliability of the RPM method can
be tested by comparing the mass inferred from this method with an independently measured $M$ value
(as mentioned in the previous paragraph).
Moreover, \cite{motta40} listed some HFQPOs simultaneously observed with Type-C QPOs 
from GRO J1655--40. They identified some of these HFQPOs as lower HFQPOs, and some as upper HFQPOs.
For the $M$ and $a/M$ values inferred by \cite{motta40}, and assuming the Type-C QPO frequencies
to be $\nu_{\rm LT}$, a radius of origin for each of these LFQPOs can be calculated for Kerr spacetime (see Sec.~\ref{s3}).
In their Fig. 5, \cite{motta40} showed that, the simultaneous lower HFQPO frequencies
match well with the $\nu_{\rm per}$ values at the corresponding radii for the same $M$ and $a/M$ values.
Similarly, the simultaneous upper HFQPO frequencies
match well with the $\nu_{\phi}$ values at the corresponding radii.
These provide a support for the RPM for QPOs.

GRO J1655--40 is currently the only BHXB, for which all the three
above mentioned $a/M$ estimation methods are available, and hence, this source
provides a unique opportunity to test the reliability of these methods by comparing the
three estimated $a/M$ values. The timing method gives 
$a/M \approx 0.286\pm0.003$ \cite{motta40},
the line spectrum method gives $a/M \approx 0.90-0.99$ \cite{Reisetal2009}, and the continuum 
spectrum method gives $a/M \approx 0.65-0.75$ (using $M \approx 6.3 M_\odot$, $D \approx 
3.2$~kpc, $i \approx 70^\circ.2$; \cite{sha}) for GRO J1655--40. Therefore, not only the $a/M$ value
inferred from the timing method is inconsistent with those inferred from the spectral methods,
but also the results from the two spectral methods are grossly inconsistent with each other.
Even if $M \approx 5.4 M_\odot$ \cite{beer} were used, which would be consistent with the
finding from the RPM method \cite{motta40}, the continuum spectrum method would give an $a/M$ range of
$0.50-0.63$. This is inconsistent with the results from both the RPM method and the
line spectrum method. These suggest that all three methods could be unreliable. If true, this will make some of
our current understandings of black holes doubtful, will deprive us of reliable $a/M$
measurement methods, and could impact the future plans of X-ray observations of BHXBs.

Can it be possible that these methods are actually reliable (as indicated by the works reported in a 
large volume of publications; e.g., \cite{ReynoldsNowak2003, Miller2007, RemillardMcClintock2006}), 
but they are missing an essential ingredient?
Here we explore an exciting possibility that the inclusion of gravitomagnetic monopole
may make the results from three methods consistent, thus suggesting that such a monopole
exists in nature. For this purpose, we allow non-zero NUT parameter values (implying 
gravitomagnetic monopole) in our calculations, by considering the KTN spacetime, instead
of the previously used Kerr spacetime. Note that the former spacetime,
having one additional parameter, i.e., the NUT parameter $n/M$, is a generalized version of the latter.
Before testing this new idea, let us first consider 
the KTN metric and derive the corresponding 
three fundamental frequencies: orbital frequency $\nu_\phi$, radial epicyclic frequency $\nu_r$
and vertical epicyclic frequency $\nu_\theta$.

\section{\label{s3}Fundamental frequencies in Kerr-Taub-NUT spacetime}
The metric of the KTN spacetime is expressed as \cite{ml}
\begin{equation}
ds^2=-\f{\d}{p^2}(dt-A d\phi)^2+\f{p^2}{\d}dr^2+p^2 d\th^2
+\f{1}{p^2}\sin^2\th(adt-Bd\phi)^2
\label{metric}
\end{equation}
with 
\begin{eqnarray}\nonumber
\d&=&r^2-2Mr+a^2-n^2, \,\,\,\,\,\,\,\,\,  p^2=r^2+(n+a\cos\th)^2,
\\
A&=&a \sin^2\th-2n\cos\th, \,\,\,\,\,\,\,\,  B=r^2+a^2+n^2,
\end{eqnarray}
where $M$ is the mass, $a/M$ is the Kerr parameter and $n/M$ is the NUT parameter.

Now, substituting the metric components ($g_{\mu\nu}$) of KTN spacetime in Eqs.~(\ref{ke}-\ref{ve})
of Appendix \ref{app},
we can obtain the three fundamental frequencies.
The orbital frequency can be written as \cite{cc2}
\begin{eqnarray}
\O_{\phi}^{\rm KTN}=2\pi\nu_{\phi}^{\rm KTN}=\pm \f{m^{\f{1}{2}}}{r^{\f{1}{2}}~(r^2+n^2) \pm a~m^{\f{1}{2}}},
\label{kktn}
\end{eqnarray}
where $m=M~(r^2-n^2)+2~n^2r$. In all the equations here, the upper sign is applicable for the
prograde orbits (which we use throughout in our paper) and the lower sign is applicable for the retrograde 
orbits.

Similarly, radial and vertical epicyclic frequencies are (which, to the best of our knowledge,
reported for the first time here):
\begin{eqnarray}\nonumber
\nu_r^{\rm KTN}&=&\f{\nu_{\phi}^{\rm KTN}}{m^{\f{1}{2}}~(r^2+n^2)}.
\left[M(r^6-n^6+15n^4r^2-15n^2r^4)-2M^2r (3r^4-2n^2r^2+3n^4)-16n^4r^3 \right.
\\
&& \left. \pm 8ar^{\f{3}{2}}m^{\f{3}{2}}
 +a^2\left\{M(n^4+6n^2r^2-3r^4)-8n^2r^3 \right\}\right]^{\f{1}{2}}
\label{rktn}
\end{eqnarray}
and 
\begin{eqnarray}\nonumber
\nu_{\th}^{\rm KTN}&=&\f{\nu_{\phi}^{\rm KTN}}{m^{\f{1}{2}}~(r^2+n^2)}.
\left[M(r^6-n^6+15n^4r^2-15n^2r^4)+2n^2r (3r^4-2n^2r^2+3n^4) + 16M^2n^2r^3 \right.
\\
&& \left. \mp 4ar^{\f{1}{2}}m^{\f{1}{2}}(n^2+Mr)(n^2+r^2)
-a^2\left\{M(n^4+6n^2r^2-3r^4)-8n^2r^3 \right\}\right]^{\f{1}{2}}
\label{vktn}
\end{eqnarray}
respectively. 

Setting the square of Eq.~(\ref{rktn}) equal to zero (i.e., $[\nu_r^{\rm KTN}]^2=0$), 
we obtain the innermost stable circular orbit (ISCO) condition as follows \cite{cc2}:
\begin{eqnarray}\nonumber
&& M(r^6-n^6+15n^4r^2-15n^2r^4)-2M^2r (3r^4-2n^2r^2+3n^4)-16n^4r^3 
\\  \nonumber
& \pm & 8ar^{\f{3}{2}}m^{\f{3}{2}} +a^2\left\{M(n^4+6n^2r^2-3r^4)
-8n^2r^3 \right\}=0.\nonumber
\\
\label{isco}
\end{eqnarray}

\subsection*{Gravitational redshift in KTN spacetime}
The gravitational redshift in the KTN spacetime is expressed as (using Eq.~\ref{Z} of Appendix \ref{app}):
\begin{eqnarray}
 Z^{\rm KTN}=\f{r^{\f{1}{2}}(r^2+n^2)+a~m^{\f{1}{2}}}{\left[(r^2+n^2)~
 \left\{r (r^2-3n^2)+M(n^2-3r^2)+2 a (m r)^{\f{1}{2}}\right\}\right]^{\f{1}{2}}}.
 \label{zktn}
\end{eqnarray}

\subsection*{SPECIAL CASES}
\subsection{Kerr spacetime ($n=0$ and $a \neq 0$)}
Now, in the Kerr spacetime ($n=0$), Eqs.~(\ref{kktn}-\ref{vktn}) reduce to \cite{ok, ka}
\begin{eqnarray}
\O_{\phi}^{\rm Kerr}=2\pi \nu_{\phi}^{\rm Kerr}=\pm \f{M^{\f{1}{2}}}{r^{\f{3}{2}} \pm a~M^{\f{1}{2}}} ,
\label{kkerr}
\end{eqnarray}

\begin{eqnarray}
\nu_{r}^{\rm Kerr}=\f{\nu_{\phi}^{\rm Kerr}}{r}.
\left[r^2 -6Mr \pm 8ar^{\f{1}{2}}M^{\f{1}{2}}-3a^2 \right]^{\f{1}{2}}
\label{rkerr}
\end{eqnarray}
and
\begin{eqnarray}
\nu_{\th}^{\rm Kerr}=\f{\nu_{\phi}^{\rm Kerr}}{r}.
\left[r^2 \mp 4ar^{\f{1}{2}}M^{\f{1}{2}}+3a^2 \right]^{\f{1}{2}}
\label{vkerr}
\end{eqnarray}
respectively.

Setting the square of Eq.~(\ref{rkerr}) equal to zero, we obtain the ISCO condition \cite{ch}:
\begin{eqnarray}
r^2 -6Mr \pm 8ar^{\f{1}{2}}M^{\f{1}{2}}-3a^2=0.
\label{iscokerr}
\end{eqnarray}

\subsection*{Gravitational redshift in Kerr spacetime}
In Kerr spacetime, gravitational redshift equation (Eq.~\ref{zktn}) reduces to
\begin{eqnarray}
 Z^{\rm Kerr}=\f{r^{\f{3}{2}}+a~M^{\f{1}{2}}}
 {r^{\f{1}{2}}\left[r^2-3 M r+2 a (M r)^{\f{1}{2}}\right]^{\f{1}{2}}}.
 \label{zkerr}
\end{eqnarray}
From the above expression, we can obtain the well-known redshift expression
in the Schwarzschild spacetime: $Z^{\rm Schwarzschild}=\left(1-\f{3M}{r}\right)^{-\f{1}{2}}$.

\subsection{NUT spacetime ($a=0$ and $n \neq 0$)}
In the case of NUT spacetime ($a=0$), Eqs.~(\ref{kktn}-\ref{vktn}) reduce to \cite{cc2}
\begin{eqnarray}
\O_{\phi}^{\rm NUT}=2\pi \nu_{\phi}^{\rm NUT}=\pm \f{m^{\f{1}{2}}}{r^{\f{1}{2}}~(r^2+n^2)} ,
\label{ktn}
\end{eqnarray}

\begin{eqnarray}\nonumber
\nu_r^{\rm NUT}=\f{\nu_{\phi}^{\rm NUT}}{m^{\f{1}{2}}~(r^2+n^2)}.
\left[M(r^6-n^6+15n^4r^2-15n^2r^4)-2M^2r (3r^4-2n^2r^2+3n^4)
-16n^4r^3\right]^{\f{1}{2}}
\\
\label{rtn}
\end{eqnarray}
and 
\begin{eqnarray}\nonumber
\nu_{\th}^{\rm NUT}&=&\f{\nu_{\phi}^{\rm NUT}}{m^{\f{1}{2}}~(r^2+n^2)}.
\left[M(r^6-n^6+15n^4r^2-15n^2r^4)+2n^2r (3r^4-2n^2r^2+3n^4)+16M^2n^2r^3 \right]^{\f{1}{2}}
\\
\label{vtn}
\end{eqnarray}
respectively. Here $m=M~(r^2-n^2)+2~n^2r$.

Setting the square of Eq.~(\ref{rtn}) equal to zero, one can obtain the ISCO condition:
\begin{eqnarray}
M(r^6-n^6+15n^4r^2-15n^2r^4)-2M^2r (3r^4-2n^2r^2+3n^4)-16n^4r^3=0.
\label{iscotn}
\end{eqnarray}

Remarkably, in general $\nu_{\phi}^{\rm NUT} \neq \nu_{\th}^{\rm NUT}$
in the NUT spacetime. This means that 
the LT precession frequency $\nu_{\rm LT}^{\rm NUT} ~(\equiv \nu_{\phi}^{\rm NUT} - \nu_{\th}^{\rm NUT}$)
does not vanish in NUT spacetime, i.e., inertial frames are dragged 
due to the presence of a {\it non-zero} NUT charge, although the spacetime is {\it non-rotating} ($a=0$). 

\subsection*{Gravitational redshift in NUT spacetime}
In NUT spacetime, the gravitational redshift equation (Eq.~\ref{zktn}) reduces to
\begin{eqnarray}
 Z^{\rm NUT}=\left[\f{r (r^2+n^2)}{r (r^2-3n^2)+M(n^2-3r^2)}\right]^{\f{1}{2}}.
 \label{znut}
\end{eqnarray}

\section{\label{s4}Exploring the possibility of non-zero NUT parameter in GRO J1655-40}

The three fundamental frequencies (Eqs. \ref{kkerr}-\ref{vkerr})
for the Kerr spacetime and for infinitesimally eccentric and tilted orbits 
were used by \cite{motta40} for $a/M$ estimation using the RPM method.
Since we use the KTN spacetime instead of the Kerr spacetime, here
we use the expressions of these frequencies (see Eqs.~\ref{kktn}--\ref{vktn})
corresponding to the KTN spacetime. One can now derive the periastron precession frequency 
$\nu_{\rm per}^{\rm KTN} (= \nu_{\phi}^{\rm KTN} - \nu_r^{\rm KTN}$) and
the Lense-Thirring (LT) precession frequency 
$\nu_{\rm LT}^{\rm KTN} (=\nu_{\phi}^{\rm KTN} - \nu_{\th}^{\rm KTN}$) 
using these three fundamental frequencies.
Besides, the condition to derive the radius of the innermost stable circular orbit 
$r_{\rm ISCO}$ and the expression of the 
gravitational redshift for the KTN spacetime
are given by Eq.~(\ref{isco}) (see also \cite{cc2}) 
and Eq.~(\ref{zktn}) respectively.

\begin{figure}[!h]
\centering{}\includegraphics[width=5.5in,height=5in,angle=0]{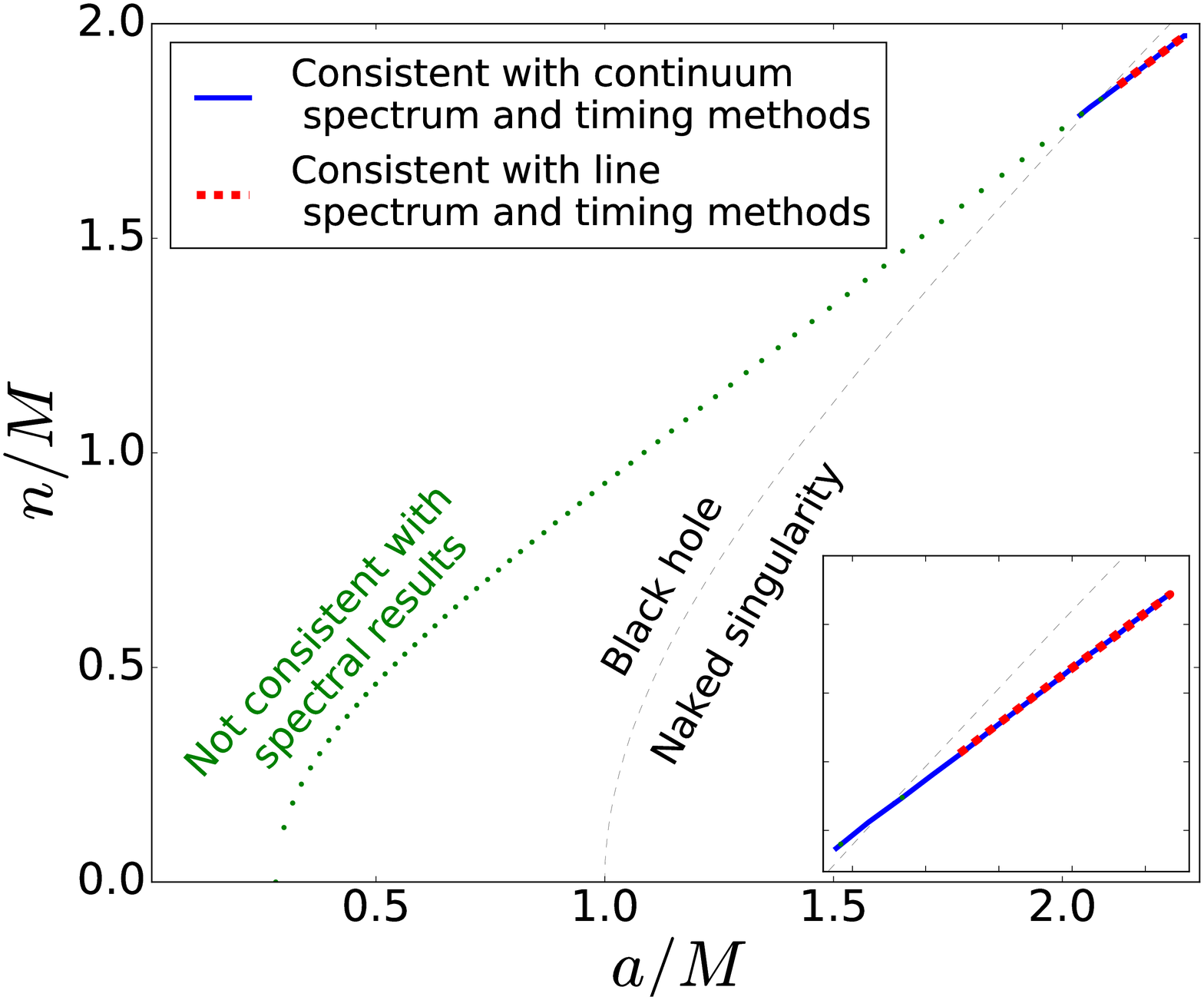}
\protect\caption{\label{fg1}The NUT parameter ($n/M$) versus
the spin parameter ($a/M$) space, which is divided into a black hole region and a 
naked singularity region (see text) by the black dashed line. The $n/M$ versus $a/M$ 
constraints for GRO J1655--40 are given by (1) the green dotted curve (using
only the RPM timing method), (2) the red dotted curve (using the RPM timing and
line spectrum methods), and (3) the blue solid curve (using the RPM timing and
continuum spectrum methods). A zoomed version of the latter two is shown in 
the inset for clarity. This figure shows that there is a range of $n/M$ and 
$a/M$ values for GRO J1655--40 allowed by all the three methods.}
\end{figure}

Now, we apply the RPM method to GRO J1655--40 using the KTN frequencies.
Following \cite{motta40}, we consider 
\begin{equation}
 \nu_{\phi}^{\rm KTN} = 440~{\rm Hz}; \,\,
 \nu_{\rm per}^{\rm KTN} = 300~{\rm Hz}; \,\,
\,\, \nu_{\rm LT}^{\rm KTN} = 17~{\rm Hz}
\label{rpmJ1655}
\end{equation}
for GRO J1655--40, and using the expressions given
in Eqs.~(\ref{kktn}--\ref{vktn}), we can solve Eqs.(\ref{rpmJ1655})
for $a/M$, $M$ and the radius $r_{\rm qpo}$ of QPO origin for a given $n/M$ value.
For $n/M = 0$, we naturally recover the $a/M$, $M$ and $r_{\rm qpo}$ values reported in \cite{motta40}. 
Now, if we increase $n/M$ from zero, $a/M$ also increases (while Eq.~\ref{rpmJ1655} is satisfied),
and hence the RPM method gives an allowed $n/M$ versus $a/M$ relation (shown by the
green dotted curve in Fig.~\ref{fg1}) for GRO J1655--40.

Note that the range of $a/M$ is $0-1$ for a Kerr black hole. For $a/M > 1$,
the radii of the horizons $r_{\pm}(=M \pm \sqrt{M^2-a^2}$) become imaginary,
and hence the collapsed object becomes a naked singularity \cite{ckj,ckp}.
However, for a KTN collapsed object the radii of the horizons are 
$M \pm \sqrt{M^2+n^2-a^2}$, and hence the condition for a naked
singularity is $a/M > \sqrt{1+(n/M)^2}$ \cite{wei}. This condition is shown by a
black dashed line in Fig.~\ref{fg1}, which divides the $n/M$ versus $a/M$
space into a black hole region and a naked singularity region for the KTN spacetime.
This figure shows that $a/M$ can easily be much higher
than $1$ for a black hole for non-zero $n/M$ values. We find that the $n/M$ versus $a/M$ curve allowed from
the RPM method for GRO J1655--40 extends into the naked singularity region (Fig.~\ref{fg1}).
Note that black holes and naked singularities could coexist in nature \cite{JoshiMalafarina2011}, and 
hence the detection of an event horizon of a collapsed object does not rule out 
the possibility of the existence of a naked singularity, and vice versa.

Let us now explore if non-zero $n/M$ values can make the $a/M$ ranges inferred from 
the RPM method and the line spectrum method consistent with each other for GRO J1655--40, 
and if so, what constraints of $a/M$ and $n/M$ can be obtained.
We do this by combining these two methods as described below.
The $a/M$ range for GRO J1655--40 was estimated to be $\approx 0.90-0.99$
using the line spectrum method \cite{Reisetal2009}. But this estimation assumed
Kerr spacetime, while we need to constrain parameters in the KTN spacetime, to allow non-zero $n/M$ values.
Therefore, using Eqs.~(\ref{iscokerr}) and (\ref{zkerr}), we calculate the gravitational redshift range ($\approx 2.70-6.08$)
from the reported $a/M$ range ($\approx 0.90-0.99$).
This gravitational redshift could be directly inferred from the extent of the red wing of the
observed broad iron line (see Sec. \ref{s2}), and itself does not depend on Kerr spacetime.
Therefore, we treat this gravitational redshift range ($\approx 2.70-6.08$) as the 
primary observational constraint, independent of the Kerr spacetime.
Using this primary constraint and assuming the KTN spacetime, i.e.,
$Z^{\rm KTN} {\rm (at}~r_{\rm ISCO}) = 2.70-6.08$ (L.H.S. is given by Eqs.~\ref{isco} and \ref{zktn}),
and using Eq.~(\ref{rpmJ1655}), we solve for $M$, $a/M$, $n/M$, $r_{\rm ISCO}/M$ and 
$r_{\rm qpo}$ (in unit of $M$). This solution gives the following constraints
for GRO J1655--40, which are consistent with both the RPM method and the line spectrum method:
$M \approx 6.76-6.83 M_\odot$, $a/M \approx 2.12-2.27$, $n/M \approx 1.86-1.97$ and $r_{\rm qpo}/M \approx 4.99-5.04$.
While the non-zero $n/M$ range implies the existence of the gravitomagnetic monopole,
the red dotted curve in Fig.~\ref{fg1} shows that this $n/M$ versus $a/M$ range implies
a naked singularity. The $M$ range is consistent with an independently measured mass
($[6.3\pm0.5] M_\odot$; \cite{Greeneetal2001}) for GRO J1655--40, which
provides a confirmation of the reliability of our method and results.

Next, we explore if the $a/M$ ranges inferred from the RPM method and the continuum spectrum method
can be consistent with each other for GRO J1655--40, if non-zero $n/M$ values are allowed.
For GRO J1655--40, the $a/M$ range estimated from the continuum spectrum method is 
$\approx 0.65-0.75$ \cite{sha}, assuming the Kerr spacetime.
Therefore, as argued in the previous paragraph, we need a primary observational 
constraint, independent of the Kerr spacetime, so that the more general 
KTN spacetime for non-zero $n/M$ values can be used.
For GRO J1655--40, we find that the quoted $a/M$ range of $\approx 0.65-0.75$ \cite{sha} was inferred from
an $r_{\rm ISCO}$ range of $\approx 29.8-34.2$~km and using $M = 6.3 M_\odot$.
As mentioned in Sec. \ref{s2},
$r_{\rm ISCO}$ could directly (i.e., independent of the Kerr spacetime) 
be inferred from the observed spectrum using the known source distance ($D$) 
and the accretion disk inclination angle ($i$) values.
Therefore, using $r_{\rm ISCO} = 29.8-34.2$~km as the 
primary constraint and assuming the KTN spacetime (e.g., using Eq.~\ref{isco}), and using Eq.~(\ref{rpmJ1655}), 
we solve for $M$, $a/M$, $n/M$, $r_{\rm ISCO}/M$ and $r_{\rm qpo}/M$.
Consequently, the following parameter constraints could be obtained:
$M \approx 6.79-6.86 M_\odot$, $a/M \approx 2.04-2.21$, $n/M \approx 1.79-1.93$
and $r_{\rm qpo}/M \approx 4.96-5.02$.
These parameter ranges are largely overlapping with those obtained from the combined RPM and line spectrum method.
We find that even for this combined RPM and continuum spectrum method,
the non-zero $n/M$ range implies the gravitomagnetic monopole,
the mass is consistent with an independently measured mass ($[6.3\pm0.5] M_\odot$; \cite{Greeneetal2001})
for GRO J1655--40, and the $n/M$ versus $a/M$
curve (the blue solid curve in Fig.~\ref{fg1}) mainly implies a naked singularity, although a black hole
is also possible.

\section{\label{s5}Other probable solutions with non-zero NUT parameter in GRO J1655--40}~
It should be noted that there is a possibility to obtain other 
solutions with the non-zero NUT parameter, and consequently, other sets of parameter constraints.
This is because the LT precession frequency can change sign as one moves outwards from the
collapsed object. This implies the same absolute value of the LT precession frequency at 
three radius values. In this section, we discuss on these other plausible solutions,
and show that those solutions are not viable.

The three simultaneous QPOs from GRO J1655--40 were used by Motta et al.
\cite{motta40} to infer the parameter values of this source using the RPM method.
These parameter values were used in Fig.~5 of their paper to make the theoretical
frequency versus radius curves (three curves for three frequencies). 
Then they collected pairs of two simultaneously observed QPOs (one is LFQPO and 
another one is HFQPO) from this source. Among the HFQPOs, they considered two as upper
HFQPOs, and rest as lower HFQPOs. Using an LFQPO frequency, and the theoretical
LT precession frequency curve (drawn using the inferred parameter values from
three simultaneous QPOs, as mentioned above), the radius of origin of the LFQPO
is calculated. Then if it is assumed that
the simultaneously observed HFQPO is originated from the same radius, the 
frequency of the HFQPO comes out to be more-or-less consistent with the theoretical frequency 
curve, as required by RPM (see Fig.~5 of Motta et al. \cite{motta40}). 
This provides a support for the RPM method to estimate $a/M$.


\begin{figure}[!h]
\centering{}\includegraphics[width=7.2in,height=4.4in,angle=0]{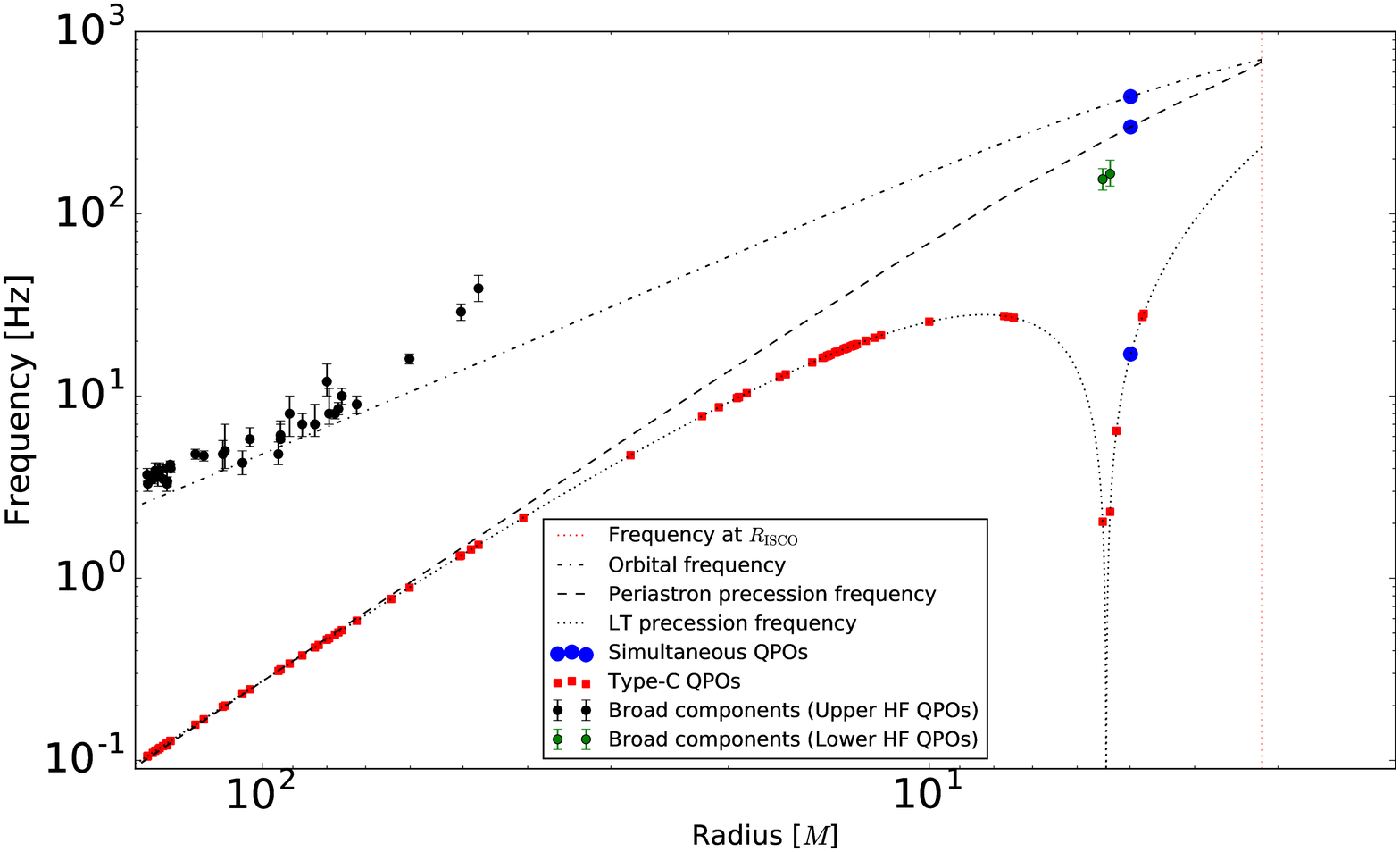}
\protect\caption{\label{fg2}LT precession frequency (dotted line), periastron precession
frequency (dashed line) and orbital frequency (dot-dashed line) as a function of the distance
$(r)$ around a KTN collapsed object as predicted by the RPM. The lines 
are drawn for $M=6.83~M_{\odot}$, $a=2.12~M$ and $n=1.86~M$.
The observed QPO frequencies (red, black and green points in the plot)
are from the Table 1 of Motta et al. \cite{motta40}.
This plot may be compared with Fig.~5 of Motta et al. \cite{motta40}
(see Sec.~\ref{s5}).}
\end{figure}

In our paper, we have considered a non-zero NUT parameter, 
which makes the results from three $a/M$ measurement methods
consistent with each other. An important point is, 
even for a non-zero NUT parameter, our results could qualitatively explain the pairs of simultaneous 
LFQPO and HFQPO by RPM (like in Fig.~5 of Motta et al. \cite{motta40}).
We show it in our Fig.~\ref{fg2}. However, a difference with Fig.~5 of 
Motta et al. \cite{motta40} is, we consider Motta et al.'s
upper HFQPOs as lower HFQPOs, and Motta et al.'s lower HFQPOs as upper HFQPOs.
Our this assumption is
not worse than Motta et al.'s assumption, because there is no independent way
to find out which HFQPOs are lower ones, and which are upper ones
(when they are not simultaneously observed). 
Note that in both figures (our Fig.~\ref{fg2} and Motta et al.'s Fig.~5), 
the data points and model curves have similar trends,
although the model curves do not quantitatively describe the data points well
either in Motta et al.'s case or in our case, possibly due to systematics related to
additional physical complexities (see Sec.\ref{s6}). Nevertheless, the qualitative matching
between the model and the data, shown in both the figures, tentatively supports 
the RPM method.

\begin{figure}[!h]
\centering{}\includegraphics[width=5.5in,height=5in,angle=0]{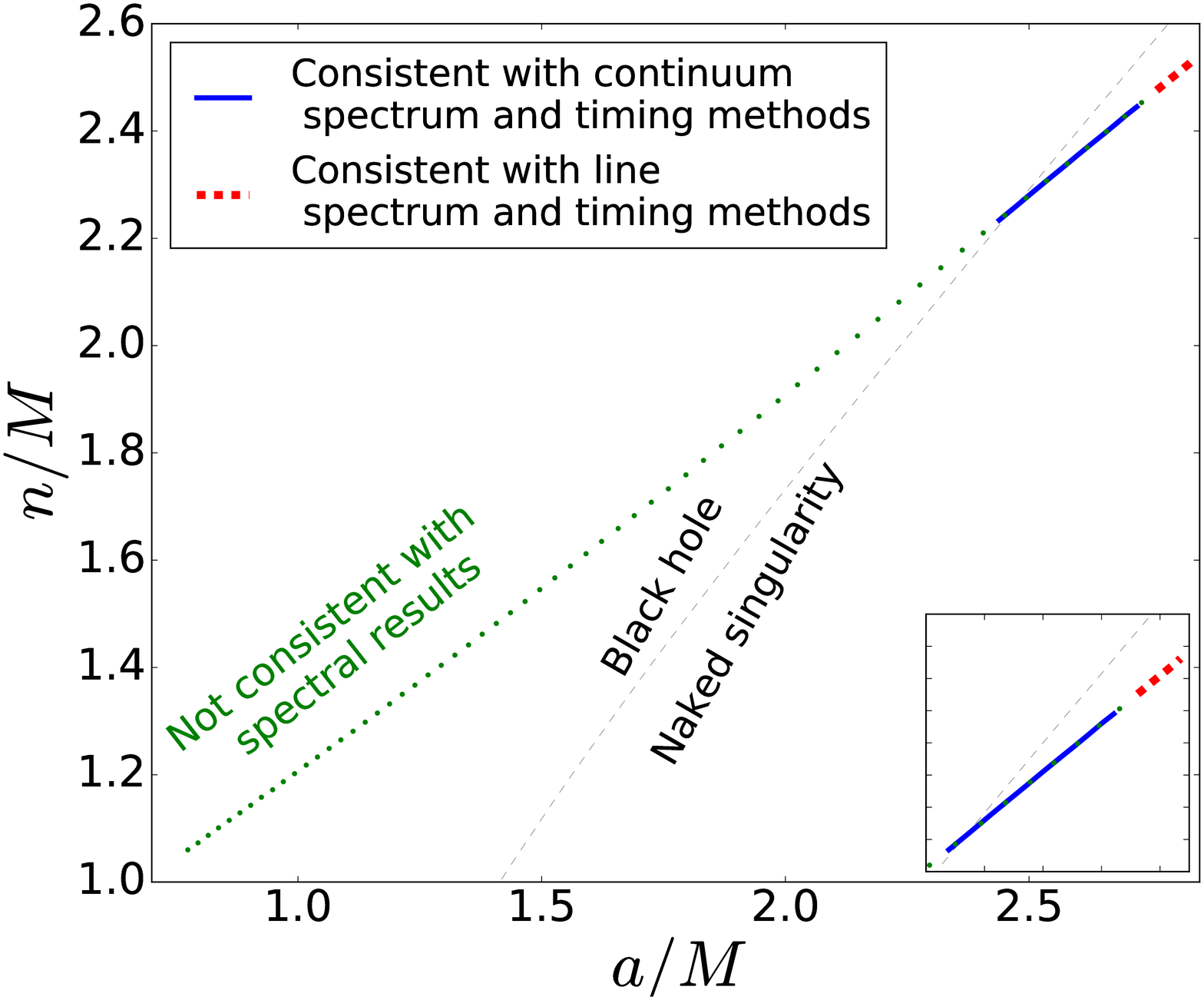}
\protect\caption{\label{fg3}The NUT parameter ($n/M$) versus
the spin parameter ($a/M$) space, which is divided into a black hole region and a 
naked singularity region by the black dashed line. The $n/M$ versus $a/M$ 
constraints for GRO J1655--40 are given by (1) the green dotted curve (using
only the RPM timing method), (2) the red dotted curve (using the RPM timing and
line spectrum methods, and (3) the blue solid curve (using the RPM timing and
continuum spectrum methods). A zoomed version of the latter two is shown in 
the inset for clarity. This figure (for the second set of the parameter values; see
Sec.~\ref{s5}) shows that there is no range of $n/M$ and 
$a/M$ values for GRO J1655--40 allowed by all the three methods, unlike Fig.~\ref{fg1}.}
\end{figure}

However, the LT precession frequency can change sign for a non-zero NUT parameter, as one
moves outwards from the collapsed object.
This implies the same absolute value of the LT precession frequency 
at three different radius values (see Fig.~\ref{fg2}). Note that we take the absolute value,
because a negative frequency only implies the opposite direction, which is
not important for our purpose. Therefore,
applying the similar method  we followed to solve Eqs.~(\ref{rpmJ1655}), two more sets of parameter 
values could be obtained from the solution of the following equations: 
\begin{equation}
 \nu_{\phi}^{\rm KTN} = 440~{\rm Hz}; \,\,
 \nu_{\rm per}^{\rm KTN} = 300~{\rm Hz}; \,\,
\,\, \nu_{\rm LT}^{\rm KTN} = - 17~{\rm Hz}.
\label{rpm2}
\end{equation}
It is clearly seen from Fig.~\ref{fg3}, that, for the second set of the parameter values,
no range of $n/M$ and $a/M$ values for GRO J1655--40 is allowed by all the three methods
(RPM, line spectrum and continuum spectrum).
Therefore, we can rule out this solution.
The third set of parameter values do not come out as real
and physical for the observed frequencies of the three simultaneous QPOs. 
Therefore, as the second solution is not
consistent with the three different spin measurement methods, and the third solution does 
not exist, we do not consider these additional sets of parameter values.
Thus, only the first solution (for the `positive' LT precession frequency), which has been
discussed in Sec.\ref{s4}, is acceptable.


\section{\label{s6}Conclusion and Discussion}

It has been shown that $a/M \approx 2.12-2.21$ and $n/M \approx 1.86-1.93$ are consistent with all the three
methods. These ranges imply that the collapsed object in GRO J1655--40 is a naked singularity
(Fig.~\ref{fg1}). Besides, the lower limit $1.86$ of $n/M$ implies the existence of the gravitomagnetic 
monopole. While this is not a direct detection of such a monopole, the indication is strong 
within our paradigm for the
following reasons. Recall that the three methods gave widely different constraints on $a/M$
($\approx 0.286\pm0.003$ \cite{motta40}; $\approx 0.90-0.99$ \cite{Reisetal2009}; $\approx 0.65-0.75$ \cite{sha}).
With only one additional parameter (i.e., the NUT parameter $n/M$), 
it might be possible to make the constraints from two of these methods consistent with each other.
We have attempted this separately for two joint methods: (1) RPM and line spectrum method, 
and (2) RPM and continuum spectrum method;
and obtained combined parameter constraints for each of these cases.
While this is not unexpected (as we have used an additional parameter $n$), the combined constraint
on $M$ being consistent with an independently measured mass value for each of the joint methods already shows the
reliability of our approach. But the main strength of our results is, we also find that the combined 
constraints on each of $n/M$ and $a/M$ from these two joint methods are largely overlapping with each other.
This cannot be a result of fine-tuning, as with just one additional 
parameter ($n$), it is not possible to fine-tune and make three different results from three independent
methods consistent with each other. Hence, the fact that we have found consistent $a/M$ and 
$n/M$ ranges from all three methods by invoking just one additional parameter ($n$) 
points to the non-zero $n$ values for GRO J1655--40, and hence suggests the existence of the gravitomagnetic monopole in nature.
This is further confirmed by the
consistent ranges of $r_{\rm qpo}/M$ ($\approx 4.99-5.02$) and $M$ ($\approx 6.79-6.83 M_\odot$)
for the methods, as well as the consistency of this $M$ range with an independently measured 
value ($[6.3\pm0.5] M_\odot$; \cite{Greeneetal2001}). This confirmation also provides a new way
to measure the NUT parameter, even when only two $a/M$ measurement methods are available for a BHXB.
It should be noted that like $a/M$, the value of $n/M$ can be different for 
different objects and a high $n/M$ value inferred for one object in this paper does not 
mean that every object will have a high $n/M$ value. The value of $n/M$ can even be
very close to zero for some objects. 
But the inferred significantly non-zero $n/M$ value for even one object could strongly 
suggest the existence of gravitomagnetic monopole in nature.
Our new technique also makes the black hole spin measurement methods more reliable.

Here we note that the `extra angular momentum' \cite{bon} makes the Taub-NUT metric singular
(coordinate singularity) at $\th=\pi$, which is a `Dirac string singularity' \cite{rs2}.
Misner \cite{mis} wanted to present an entirely nonsingular
cosmological model (homogeneous and anisotropic) with the Taub-NUT metric,
which contains the closed spacelike
hypersurfaces (but no matter), and this made this metric singularity-free.
Ramaswamy and Sen \cite{rs,rs2} pointed out that the presence of NUT parameter
requires that either the Taub-NUT metric can be singular (not the curvature) or the spacetime
contains closed timelike curves.
Since, in this paper, we have also included a possibility of the KTN `naked singularity',
we do not require the `singularity-free spacetime' to interpret our results.
This means that the `closed timelike curves' are not required for our interpretation.

Note that we have not fit the observed spectra
with KTN spectral models, because such models are not currently available. Instead,
for the purpose of $a/M$ estimation, we have used $r_{\rm ISCO}$
and the gravitational redshift at $r_{\rm ISCO}$ as proxies for the details of 
continuum spectrum and line spectrum respectively.
As argued in this paper, the use of these proxies is reasonable,
although such a use can introduce some systematics in the inferred parameter ranges. 
However, given that the inferred $n/M$ range ($\approx 1.86-1.93$) is significantly away from $n/M = 0$
(see Fig.~\ref{fg1}), the inferred non-zero $n/M$ values cannot be caused by these systematics.
Besides, $n/M = 0$ gives three widely different $a/M$ ranges from three different methods
for GRO J1655--40, as discussed earlier. 
Therefore, this paper presents the first significant observational indication of
the gravitomagnetic monopole, which, even though is not a direct detection,
can have an exciting impact on fundamental physics and astrophysics.
However, although the allowed $n/M$ versus $a/M$ range is in the naked singularity
region (Fig.~\ref{fg1}), it is close to the border of the black hole region, and hence the
indication of a naked singularity is only suggestive.

Finally, note that our inference of a non-zero NUT parameter could be correct for 
our assumption, i.e., the three existing methods of black hole spin measurements 
are reliable. However, one or more of these methods may not be entirely reliable 
due to additional physical complexities. Some of these complexities may be due to 
the following reasons (e.g., \cite{kraw} discusses how difficult it is to test the 
Kerr metric with X-ray observations). (1) The continuum X-ray spectrum method assumes that
the thin disk emission can be fully separated from emissions
from other X-ray components, which may not be correct. (2) Spectral methods also
assume that the black hole's spin
is aligned with the inner disk angular momentum vector, which is not
necessarily true \cite{ChakrabortyBhattacharyya2017}. (3) The
relativistic precession model assumes that particles in the accretion
disk travel on exact geodesic orbits, and neglects important physics
such as radiation physics, viscosity, magnetic fields that could
affect the motion of material in the disk. While there is a possibility that 
such systematic uncertainties could explain the three different ranges of spin values obtained
from three methods for $n = 0$, such a level of unreliability of
the methods would make many of the current black hole studies
doubtful and could impact the plans of X-ray observations of BHXBs
with future space missions (Sec.\ref{s2}).
\\

\appendix

\begin{appendix}

 \section{\label{app}Fundamental frequencies in a general stationary and axisymmetric spacetime}
Let us consider a general stationary and axisymmetric spacetime as
\begin{eqnarray}
 ds^2=g_{tt}dt^2+2g_{t\phi}d\phi dt+g_{\phi\phi}d\phi^2+g_{rr}dr^2+g_{\th\th}d\th^2,
 \label{met}
\end{eqnarray}
where $g_{\mu\nu}=g_{\mu\nu}(r, \th)$. In this spacetime, 
the proper angular momentum ($l$) of a test particle can be defined as: 
\begin{eqnarray}
 l=-\frac{g_{t\phi}+\Omega_{\phi} g_{\phi\phi}}{g_{tt}+\Omega_{\phi} g_{t\phi}},
\end{eqnarray}
where, $\Omega_{\phi}$ is the orbital frequency of the test particle. $\Omega_{\phi}$
is defined as \cite{don}
\begin{eqnarray}
\Omega_{\phi} \equiv  2\pi\nu_{\phi} =\frac{d\phi/d\tau}{dt/d\tau}=\frac{d\phi}{dt}
 =\frac{-g_{t\phi}'\pm \sqrt{g_{t\phi}'^2-g_{tt}'g_{\phi\phi}'}}{g_{\phi\phi}'}\mid_{r={\rm constant}, \th \rightarrow \pi/2}, 
 \label{ke}
 \end{eqnarray}
where the prime denotes the partial differentiation with respect to $r$.
The general expressions for calculating the radial ($\nu_r$) and vertical 
($\nu_{\th}$) epicyclic frequencies are \cite{don}
\begin{eqnarray}
 \nu_r^2
=\f{(g_{tt}+\O_{\phi}g_{t\phi})^2}{2 (2\pi)^2~g_{rr}}\left[\p_r^2\left({g_{\phi\phi}}/{Y}\right)
 +2l~\p_r^2\left({g_{t\phi}}/{Y}\right)+l^2~\p_r^2\left({g_{tt}}/{Y}\right) \right]|_{r={\rm constant}, \th \rightarrow \pi/2}\nonumber
 \\
 \label{re}
\end{eqnarray}
and
\begin{eqnarray}
\nu_{\th}^2
=\f{(g_{tt}+\O_{\phi}g_{t\phi})^2}{2 (2\pi)^2~g_{\th\th}}\left[\p_{\th}^2\left({g_{\phi\phi}}/{Y}\right)
 +2l~\p_{\th}^2\left({g_{t\phi}}/{Y}\right)+l^2~\p_{\th}^2\left({g_{tt}}/{Y}\right) \right]|_{r={\rm constant}, \th \rightarrow \pi/2}\nonumber
 \\
 \label{ve}
\end{eqnarray}
respectively, and $Y$ is defined as
\begin{eqnarray}
 Y=g_{tt}g_{\phi\phi}-g_{t\phi}^2 .
\end{eqnarray}

\subsection*{Gravitational redshift}
The general expression of gravitational redshift ($z$) in an axisymmetric and stationary
spacetime can be obtained from \cite{mtw,lum}
\begin{eqnarray}
 Z=1+z=(-g_{tt}-2\O_{\phi} g_{t\phi}-\O_{\phi}^2g_{\phi\phi})^{-\f{1}{2}}.
 \label{Z}
\end{eqnarray} 
Now, substituting the expressions of metric components ($g_{\mu\nu}$) and the orbital frequency ($\O_{\phi}$) 
in Eq.~(\ref{Z}), one can derive the expression of $Z$ of a particular  
axisymmetric and stationary spacetime, i.e., KTN, Kerr, NUT, etc.
We discuss these in Sec. \ref{s3}.
\end{appendix}
\\

{\bf Acknowledgements :} We thank the referee for the constructive comments and 
valuable suggestions. One of us (C. C.) also thanks T. Baug for some useful discussions.
C. C. gratefully acknowledges support from the National Natural Science Foundation 
of China (NSFC), Grant No. : 11750110410.


\begin{thebibliography}{}
\bibitem{dirac} P. A. M. Dirac, {\it Proc. R. Soc. Lond. A} {\bf 133}, 60 (1931)
\bibitem{saha} M. N. Saha, {\it Indian Journal of Physics} {\bf X}, 141 (1936)
\bibitem{zee} A. Zee, {\it Phys. Rev. Lett.} {\bf 55}, 2379 (1985)
\bibitem{nut} E. Newman, L. Tamburino, T. Unti, {\it J. Math. Phys.} {\bf 4}, 915 (1963)
\bibitem{ahm} V. Kagramanova, B. Ahmedov, {\it Gen. Relativ. Gravit.} {\bf 38}, 823 (2006)
\bibitem{bini} D. Bini, C. Cherubini, R. T. Jantzen, B. Mashhoon, {\it Class. Quantum Grav.} {\bf 20}, 457 (2003)
\bibitem{rs2} S. Ramaswamy, A. Sen, {\it Phys. Rev. Lett.} {\bf 57}, 1088 (1986)
\bibitem{dn} M. Demianski, E.T. Newman, {\it Bull. Acad. Pol. Sci., 
Ser. Sci. Math. Astron. Phys.} {\bf 14}, 653 (1966)
\bibitem{bon} W. B. Bonnor, {\it Proc. Camb. Phil. Soc.} {\bf 66}, 145 (1969)
\bibitem{dow} J. S. Dowker, {\it Gen. Rel. Grav.} {\bf 5}, 603 (1974) 
\bibitem{rs} S. Ramaswamy, A. Sen, {\it J. Math. Phys. (N.Y.)} {\bf 22}, 2612 (1981)
\bibitem{mp} M. Mueller, M. J. Perry, {\it Class. Quantum Grav.} {\bf 3}, 65 (1986)
\bibitem{lnbl} D. Lynden-Bell, M. Nouri-Zonoz, {\it Rev. Mod. Phys.} {\bf 70}, 427 (1998)
\bibitem{kag} V. Kagramanova, J. Kunz, E. Hackmann, C. L\"ammerzahl, {\it Phys. Rev.} {\bf D81}, 124044 (2010)
\bibitem{kerr} R. Kerr, {\it Phys. Rev. Lett.} {\bf 11}, 237 (1963)
\bibitem{ml} J. G. Miller, {\it J. Math. Phys.} {\bf 14}, 486 (1973)
\bibitem{ckp} C. Chakraborty, P. Kocherlakota, M. Patil, S. Bhattacharyya, 
P. S. Joshi, A. Kr\'olak, {\it Phys. Rev.} {\bf D 95}, 084024 (2017)
\bibitem{Miller2007} J. M. Miller, {\it ARAA} {\bf 45}, 441 (2007)
\bibitem{RemillardMcClintock2006} R. A., Remillard, J. E. McClintock, {\it ARAA} {\bf 44}, 49 (2006)
\bibitem{bs14} T. M. Belloni, L. Stella, {\it Space Science Reviews} {\bf 183}, 43 (2014)
\bibitem{motta40} S. E. Motta, T. M. Belloni, L. Stella, T. Munoz-Darias,
R. Fender, {\it Mon. Not. R. Astron. Soc.} {\bf 437}, 2554 (2014)
\bibitem{ReynoldsNowak2003} C. S. Reynolds, M. A. Nowak, {\it Phys. Rep.} {\bf 377}, 389 (2003)
\bibitem{Reisetal2009} R. C. Reis, A. C. Fabian, R. R. Ross, J. M. Miller, {\it Mon. Not. R. Astron. Soc.} {\bf 395}, 1257 (2009)
\bibitem{FragosMcClintock2015} T. Fragos, J. E. McClintock, {\it Astrophys J.} {\bf 800}, 17 (2015)
\bibitem{Bellonietal2012} T. M. Belloni, A. Sanna, M. M\'endez, {\it Mon. Not. R. Astron. Soc.} {\bf 426}, 1701 (2012)
\bibitem{StellaVietri1998} L. Stella, M. Vietri, {\it Astrophys J.}, {\bf 492}, L59 (1998)
\bibitem{StellaVietri1999} L. Stella, M. Vietri, {\it Phys. Rev. Lett.} {\bf 82}, 17 (1999)
\bibitem{Greeneetal2001} J. Greene, C. D. Bailyn, J. A. Orosz, {\it Astrophys. J.} {\bf 554}, 1290 (2001)
\bibitem{beer}  M. E. Beer, P. Podsiadlowski, {\it Mon. Not. R. Astron. Soc.} {\bf 331}, 351 (2002)
\bibitem{in16} A. Ingram, M. van der Klis, M. Middleton, C. Done, D. Altamirano, 
L. Heil, P. Uttley, M. Axelsson, {\it Mon. Not. R. Astron. Soc.} {\bf 461}, 1967 (2016)
\bibitem{MillerHoman2005} J. M. Miller, J. Homan, {\it Astrophys J.} {\bf 618}, L107 (2005)
\bibitem{Schnittmanetal2006} J. D. Schnittman, J. Homan, J. M. Miller, {\it Astrophys J.} {\bf 642}, 420 (2006)
\bibitem{ChakrabortyBhattacharyya2017} C. Chakraborty, S. Bhattacharyya, 
{\it Mon. Not. R. Astron. Soc.} {\bf 469}, 3062 (2017)
\bibitem{sha} R. Shafee, J. E. McClintock, R. Narayan, S. W. Davis, L.-X. Li, 
R. A. Remillard, {\it Astrophys. J.} {\bf 636}, L113 (2006)
\bibitem{cc2} C. Chakraborty, {\it Eur. Phys. J.} {\bf C74}, 2759 (2014)
\bibitem{ok} A. T. Okazaki, S. Kato, J. Fukue, {\it PASJ} {\bf 39}, 457 (1987)
\bibitem{ka} S. Kato, {\it PASJ} {\bf 42}, 99 (1990)
\bibitem{ch} S. Chandrasekhar, {\it The Mathematical 
Theory of Black Holes}, Oxford (1992)
\bibitem{ckj} C. Chakraborty, P. Kocherlakota, P. S. Joshi, {\it Phys. Rev.} {\bf D 95}, 044006 (2017)
\bibitem{wei} S-W. Wei, Y-X. Liu, C-E Fu, K. Yang, {\it JCAP} 10 (2012) 053
\bibitem{JoshiMalafarina2011} P. S. Joshi, D. Malafarina, {\it IJMPD} {\bf 20}, 2641 (2011)
\bibitem{mis} C. W. Misner, {\it J. Math. Phys.} {\bf 4}, 924 (1963) 
\bibitem{kraw} H. Krawczynski, {\it Gen. Relativ. Gravit.} {\bf 50}, 100 (2018) 
\bibitem{don} D. D. Doneva, S. S. Yazadjiev, N. Stergioulas, K. D. Kokkotas,
T. M. Athanasiadis, {\it Phys. Rev.} {\bf D 90}, {044004} (2014)
\bibitem{mtw} C. W. Misner, K. S. Thorne, J. A. Wheeler, 
{\it Gravitation}, W. H Freeman $\&$ Company, (1973)
\bibitem{lum} J.-P. Luminet, {\it Astron. and Astrophys.} {\bf 75}, 228 (1979)
\end{thebibliography}
\end{document}